\documentclass[journal]{IEEEtran}
\usepackage{subfigure}
\usepackage{graphicx}
\usepackage{epsfig,latexsym,amsmath,graphics}
\usepackage{amssymb,dsfont,amsthm}
\usepackage{cite,color}
\usepackage{url}
\usepackage{booktabs}
\usepackage{algorithm, algorithmicx}
\usepackage{algpseudocode}
\usepackage{stfloats}
\usepackage{verbatim}
\usepackage{bm}
\usepackage{stfloats}
\usepackage{multirow}
\usepackage{ulem}
\usepackage{cases}

\begin{document}
%
\title{The Analysis and Performance of LODC-OFDM Signal in Nonlinear Rydberg Atomic Sensor}
%
%
%
%

\author{Hao Wu, Xinyuan Yao, Rui Ni and Chen Gong
	\thanks{This work was supported by National Natural Science Foundation of China	under Grant 62331024 and 62171428.}
	\thanks{Hao Wu, Xinyuan Yao, Chen Gong are with the School of Information Science and Technology in University of Science and Technology of China, Email address: \{wuhao0719, yxy200127\}@mail.ustc.edu.cn, cgong821@ustc.edu.cn.

    Rui Ni is with Huawei Technology, Email address: raney.nirui@huawei.com.}
    }

\maketitle

\begin{abstract}
Rydberg atomic sensors have been seen as novel radio frequency (RF) measurements and the high sensitivity to a large range of frequencies makes it attractive for communications reception. However, the signal sensing process in Rydberg system involves sequential transduction from electromagnetic waves to optical signals and finally to electrical signals. The unipolar characteristic of the optical interface inherently restricts conventional OFDM reception. Therefore, adopting unipolar OFDM schemes, inspired by optical communication systems, becomes essential for compatible signal transmission. In this work, we investigate the amplitude modulation-to-amplitude modulation (AM-AM) characteristics of Rydberg atomic sensors, establishing an empirical approximation function. Building on the direct current-biased optical orthogonal frequency division multiplexing (DCO-OFDM) framework, we propose a novel local oscillator direct current-biased OFDM (LODC-OFDM) scheme specifically optimized for Rydberg-based sensing, effectively addressing the broadband OFDM reception challenge. Then, we adopt Bussgang theorem to analyze the nonlinear distortion of LODC-OFDM signals and the results in closed-form solutions are derived for AM/AM curves approximated by Taylor series expansion and for the ideal pre-distortion case. In real experiments, the experimental and theoretical results fit well.

\end{abstract}

\section{Introduction}
 
Rydberg atomic sensors represent a novel class of electric field measurement devices that enable direct SI-traceable and self-calibrated detection\cite{anderson2021self,song2017quantum}. These sensors exploit the exceptional electric field sensitivity of Rydberg atoms in highly excited states, making them particularly suitable for developing atom-based receivers for communication signal detection. Rydberg atom-based electric field sensing has achieved remarkable progress in both field strength and frequency coverage. It has been employed for field strengths measurement from nV/m\cite{jing2020atomic,gordon2019weak} to kV/m scale\cite{anderson2016optical,paradis2019atomic,holloway2014broadband} covering frequency from below 1 kHz\cite{jau2020vapor} to GHz, THz and PHz\cite{downes2020full,wade2017real,wang2023theoretical}.

Currently, Rydberg atom-based microwave communication has been achieved in the many bands and modulation methods\cite{artusio2022modern,holloway2020multiple,simons2019embedding,liu2023electric,meyer2018digital,liu2022deep,zhang2022rydberg,zhang2024image}. Especially in work \cite{meyer2018digital}, the quadrature amplitude modulation (QAM) with 1 MHz bandwidth has been achieved. And in work \cite{liu2022deep}, deep learning provides a solution for Rydberg atom broadband communication which allows for direct decoding of frequency division multiplexed signals. However, no experimental demonstration of broadband orthogonal frequency division multiplexing (OFDM) signal reception has been reported for Rydberg-based sensors, despite the widespread adoption of OFDM in conventional microwave communication systems. This research gap stems from fundamental challenges imposed by the unique transduction characteristics of Rydberg atoms. The reception of atomic sensors involves sequential transduction from electromagnetic waves to optical signals and finally to electrical signals. The unipolar characteristic of the optical interface inherently restricts conventional OFDM reception. Therefore, adopting unipolar OFDM schemes, inspired by optical communication systems, becomes essential for signal transmission.

Among various optical orthogonal frequency-division multiplexing (O-OFDM) schemes \cite{Jean2009OFDM}, direct current-biased optical orthogonal frequency division multiplexing (DCO-OFDM) and asymmetrically clipped optical orthogonal frequency division multiplexing (ACO-OFDM) represent two predominant implementations, each exhibiting distinct characteristics in signal generation and power efficiency. The DCO-OFDM scheme employs a sufficiently large direct current (DC) bias to convert bipolar OFDM signals into unipolar waveforms suitable for optical intensity modulation, albeit at the cost of significant power wastage and consequently reduced system energy efficiency. In contrast, the ACO-OFDM approach utilizes only odd subcarriers for data modulation and generates unipolar signals through asymmetric clipping that eliminates the negative signal components. This method offers improved energy efficiency by eliminating the need for DC biasing, but achieves this advantage at the expense of halved spectral utilization efficiency due to its reduced subcarrier allocation.

Consider the intrinsic high peak-to-average power ratio of OFDM signals and the sensitivity to phase noise and nonlinear distortion, a investigation of Rydberg atomic nonlinearities on OFDM signal is imperative. In traditional radio freqency (RF) antenna scenarios, the effects of the non-linearity and the phase noise on the OFDM system performance have been investigated in the past\cite{2002M,1999Impact,Banelli2000Theoretical}. In optical communication scenarios, unipolarity constraint of intensity modulation and the limited dynamic range of LED to guarantee the waveform non negativity result in the double-sided clipping distortion on the signal \cite{2020Optimum,2015Offset,2016PAPR}. The sensing mechanism of Rydberg sensors involves both electromagnetic and optical fields, necessitating a comprehensive analysis of OFDM signal effects that must account for both nonlinear distortion and clipping-induced impairments.

In this work, we develop a local oscillator direct-current-biased OFDM (LODC-OFDM) architecture specifically tailored for Rydberg sensing, drawing inspiration from optical communication systems while addressing the unique unipolar constraints of Rydberg atomic receivers. Accounting for OFDM signals' characteristically high peak-to-average power ratio and their sensitivity to nonlinear distortion, we investigate the amplitude modulation-to-amplitude modulation (AM-AM) conversion characteristics of  Rydberg atomic sensors based on a four level model. Our analysis yields an empirical transfer function that captures the nonlinear AM-AM response characteristics. Then, we adopt Bussgang theorem to analyze the nonlinear distortion of LODC-OFDM signals and the results in closed-form solutions are derived for AM/AM curves approximated by Taylor series expansion and for the ideal pre-distortion case. In real experiments, the experimental and theoretical results fit well.

The remainder of this paper is organized as follows. In Section \ref{T2}, we introduce a four-level Rydberg atomic model and its associated detection system. From this framework, we derive an approximate analytical expression for the AM-AM transfer function, with experimental results validating the formulation's accuracy. In Section \ref{T3}, we propose LODC-OFDM scheme and analyze its performance adopting Bussgang theorem. In Section \ref{T4}, we show the experimental and numerical results including bit error rate and channel capacity. Finally, Section \ref{T5} concludes this work.

%
%
%
%


\section{Physical Model and AM-AM Function}\label{T2}
\subsection{Four-level Rydberg System Diagram}\label{T21}
\begin{figure*}
	\centering
	\includegraphics[width=0.75\textwidth]{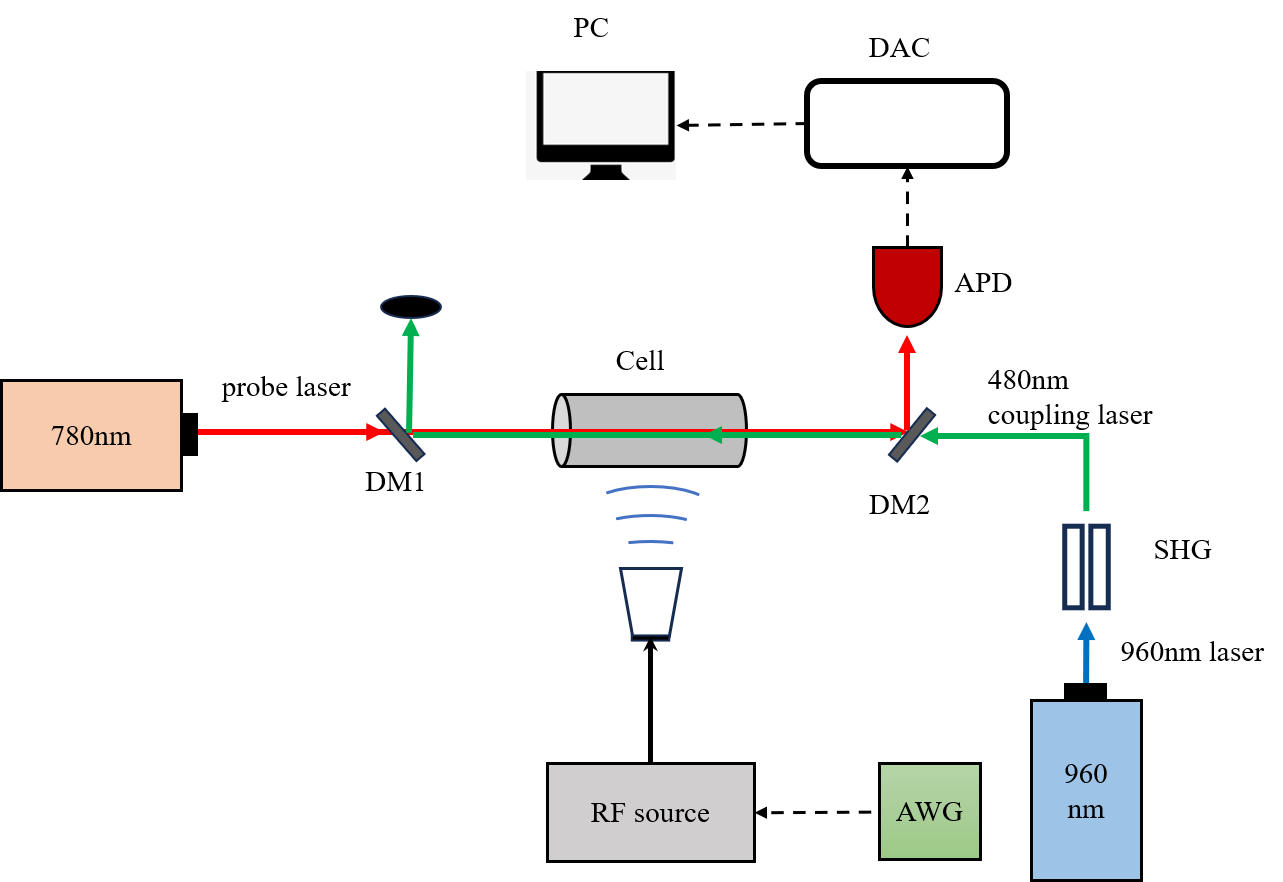}
	\caption{The diagram for the four-level scheme.}
	\label{systemdiagram}
\end{figure*}

A schematic diagram of the detection system based on Rydberg atomic sensor is shown in Fig. \ref{systemdiagram}. We employed a 780 nm laser as the probe laser and a 480 nm as the coupling laser. These lasers propagate in opposite directions and excite part of atoms inside the atomic cell to Rydberg states. Eventually, the probe and coupling laser are separated by two dichroic mirrors (DM). The output probe laser, passing through atomic cell, is received by an avalanche photodiode (APD). As the RF signals, controlled by the arbitrary waveform generator (AWG), propagate through the Rb atom vapor, the transmittance of Rydberg atomic gas would be effected. The corresponding data is collected by a data acquisition card (DAC)  and stored in personal computer (PC) for offline symbol detection. In this work, we assume that the probe laser, coupling laser, and RF field share the same polarization\cite{sedlacek2013atom}.

\subsection{Theoretical Physical Model}\label{T22}

\begin{figure}[htbp]
	\centering
	\includegraphics[width=0.45\textwidth]{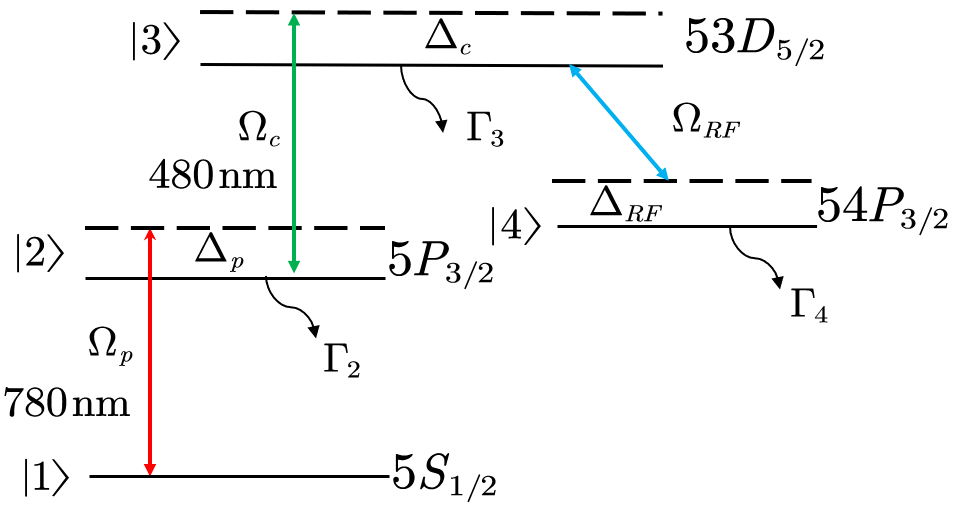}
	\caption{The diagram for the four-level scheme.}
	\label{diagram}
\end{figure}

The typical Rydberg four-level ladder system is shown in Fig. \ref{diagram}. The probe and coupling laser beams excite the atoms from state $\left| 1 \right> $ to state $\left| 2 \right> $ and from state $\left| 2 \right> $ to state $\left| 3 \right>$, respectively. The RF field couples Rydberg state $\left| 3 \right>$ and state $\left| 4 \right> $. Let $\Delta _p$, $\Delta _c$, and $\Delta _{RF}$ denote the detunings for the probe laser, couple
laser and RF electric field, respectively. Let $\Omega _p$, $\Omega _c$, and $\Omega _{RF}$ denote Rabi frequencies associated with the probe laser, couple laser and RF electric field, respectively. We have
\begin{equation}
	\begin{aligned}
		\Omega _{p}&=|E_{p}|\frac{\mu _{p}}{\hbar },\\
		 \Omega _{c}&=|E_{c}|\frac{\mu _{c}}{\hbar },\\
		  \Omega _{RF}&=|E_{RF}|\frac{\mu _{RF}}{\hbar },
	\end{aligned}
\end{equation}	
where $|E_{p}|$, $|E_{c}|$ and $|E_{RF}|$ are the magnitudes of the electric-field of the probe laser, coupling laser, and RF source, respectively. Let $\mu _{p}$, $\mu _{c}$ and $\mu _{RF}$ denote the atomic dipole moments corresponding to $\left| 1 \right> -\left| 2 \right> $, $\left| 2 \right> -\left| 3 \right> $ and $\left| 3 \right> -\left| 4 \right>$, respectively. Let $\hbar$ denotes reduced planck constant. Assume that the decay rates are $\Gamma_{2}$ for state $\left| 2 \right>$, $\Gamma_{3}$ for state $\left| 3 \right>$, and $\Gamma_{4}$ for state $\left| 4 \right>$\cite{holloway2017electric}.

The Lindblad master equation of above atomic system is given as follows\cite{auzinsh2010optically}:
\begin{equation}\label{Lindblad master equation}
	\begin{aligned}
		\mathbf{\bm{\dot\rho}}=\frac{\partial \bm{\rho }}{\partial t}=-\frac{i}{{\hbar }}\left[ \bm{H},\bm{\rho } \right] +\bm{\mathcal{L}},
	\end{aligned}
\end{equation}
where $\bm\rho$ is the density matrix of the atomic ensemble and $\bm{\mathcal{L}}$ denotes Lindblad operator.
The interaction between RF signals and Rb atoms affects component $\rho_{21}$ of the $4\times4$ matrix $\mathbf{\bm{\rho}}$. Element $\rho_{21}$ in the second row and first column of matrix $\bm{\rho}$ is related to the transmittance of atom vapor cell, which finally affects the output probe laser power. The corresponding Hamiltonian in the rotating wave approximation frame is given by

\begin{small}
\begin{equation}
	\begin{aligned}
\boldsymbol{H}=\frac{{\hbar }}{2}\left[ \begin{matrix}
	0&		\Omega _p&		0&		0\\
	\Omega _p&		-2\Delta _p&		\Omega _c&		0\\
	0&		\Omega _c&		-2\left( \Delta _p+\Delta _c \right)&		\Omega _{RF}\\
	0&		0&		\Omega _{RF}&		-2\left( \Delta _p+\Delta _c+\Delta _{RF} \right)\\
\end{matrix} \right] .
	\end{aligned}
\end{equation}
\end{small}

In the case of thermal atom model, Doppler effect caused by the atomic thermal motion. The Doppler averaged density matrix element $\rho_{21D}$ can be expressed as\cite{holloway2017electric}
\begin{equation}
	\begin{aligned}
		\rho _{21D}=\frac{1}{\sqrt{\pi}u}\int_{3u}^{3u}{\rho _{21}\left( \Delta _{p}^{'},\Delta _{c}^{'} \right) e^{-\frac{v^2}{u^2}}dv},
	\end{aligned}
\end{equation}
where $u =\sqrt{k_BT/m}$, $m$ is the atom's mass, $k_B$ is the Boltzmann constant, and $T$ is the thermodynamic temperature. At room temperature, $T = 303.15 \text{K}$, due to atomic thermal motion, the probe and coupling light detuning $\Delta _{p}^{'}$ and $\Delta _{c}^{'}$ should be modified by the following

\begin{equation}
	\begin{aligned}
		\Delta _{p}^{'}=\Delta _p-\frac{2\pi}{\lambda _p}v,
		\\
		\Delta _{c}^{'}=\Delta _c+\frac{2\pi}{\lambda _c}v.
	\end{aligned}
\end{equation}

The relationship between the atomic medium's transmittance $T_p$ and the Doppler-modified density matrix element $\rho_{21D}$ is expressed as
\begin{equation}
	\begin{aligned}
		T_p=e^{\frac{2N_0\mu _{p}^{2}kL}{\epsilon {\hbar }\Omega _p}\Im \left( \rho_{21D} \right)},
	\end{aligned}
\end{equation}
where $N_0$ is the total density of atoms and $\epsilon$ is the permittivity in vacuum. Let $L$ denote the length of atomic cell and $k=2\pi/\lambda_p$ denote the wave vector of the probe laser. Let $\Im (\rho_{21D})$ characterize the imaginary part of $\rho_{21D}$.

\subsection{Approximation of the AM-AM Transfer Function}\label{T23}

Derivation of the AM-AM transfer function requires establishing the functional relationship between the applied electric field strength and the atomic medium's transmittance. However, obtaining cavity transmittance $T_p$ from full density-matrix steady-state solutions is computationally intensive. Thus, we implement justified two simplifications that preserve experimental validity:
\begin{itemize}
	\item[$\bullet$] Simplification A: Zero detuning for both optical and microwave carriers,
\end{itemize}
\begin{itemize}
	\item[$\bullet$] Simplification B: Neglect of Doppler effect,
\end{itemize}

In simplifying A, we neglect the effects of laser detuning and microwave detuning on the AM-AM function. The former can be easily obtained by setting the laser parameters at the resonant frequency of the energy level, while the latter assumes that the Rydberg atomic sensor has the same, linearly additive response to all subcarriers at different frequencies. This approximation condition requires that the bandwidth of the OFDM frequency must be smaller than the inherent bandwidth of the Rydberg system. For the chosen Rydberg states in our experimental configuration, the system maintains optimal measurement fidelity when the signal bandwidth remains below 1 MHz.

In Approximation B, we neglect of Doppler effect in the thermal atom model. And in order to maintain consistency between the simplified (Doppler-free) model and experimental observations, we empirically calibrate the AM-AM transfer function of the simplified model through systematic fitting to experimental measurements.

In summary, accurate determination of the AM-AM transfer function for Rydberg systems requires empirical correction of the simplified model's transfer function through experimental data.

\subsubsection{AM-AM Function under Doppler-free model}

Under resonant conditions, the detuning frequencies of optical fields and microwave field are zero, $\Delta_{p}=\Delta_{c}=\Delta_{RF}=0$. Consider $\Gamma_{2}\gg  \Gamma_{3}> \Gamma_{4}$ and assuming $\Gamma_{3}= \Gamma_{4}=0$, the steady-state solution of Eq. (\ref{Lindblad master equation}) ($\frac{d\boldsymbol{\rho }}{dt}=0$) is given by

\begin{equation}
	\begin{aligned}
		\rho _{21}&=-j\frac{\Gamma _2\Omega _p\Omega _{RF}^{2}}{2\Omega _{p}^{4}+2\Omega _{c}^{2}\Omega _{p}^{2}+\left( \Gamma _{2}^{2}+2\Omega _{p}^{2} \right) \Omega _{RF}^{2}}
		\\
		&=-j\frac{\frac{\Gamma _2\Omega _p}{\Gamma _{2}^{2}+2\Omega _{p}^{2}}\Omega _{RF}^{2}}{\frac{2\Omega _{p}^{4}+2\Omega _{c}^{2}\Omega _{p}^{2}}{\Gamma _{2}^{2}+2\Omega _{p}^{2}}+\Omega _{RF}^{2}},
	\end{aligned}
\end{equation}
where $j$ denotes the imaginary unit. Then, under Doppler-free model, the relationship between transmittance $T_p$ and $\rho_{21}$ is given by
\begin{equation}
	\begin{aligned}
		T_p&=e^{\frac{2N_0\mu _{p}^{2}k_pL}{\epsilon {\hbar }\Omega _p}\Im \left( \rho _{21} \right)}
		\\
		&=e^{-\frac{\tilde{a}\Omega _{RF}^{2}}{\tilde{b}+\Omega _{RF}^{2}}},
	\end{aligned}
\end{equation}
where $\tilde{a}=\frac{2N_0\mu _{p}^{2}k_pL}{\epsilon {\hbar }\Omega _p}\cdot \frac{\Gamma _2\Omega _p}{\Gamma _{2}^{2}+2\Omega _{p}^{2}}$ and $\tilde{b}=\frac{2\Omega _{p}^{4}+2\Omega _{c}^{2}\Omega _{p}^{2}}{\Gamma _{2}^{2}+2\Omega _{p}^{2}}$. 

\subsubsection{Experimental Calibration}
The APD receives probe laser passing through the atomic gas and converts it into the electrical signal, which is proportional to transmittance $T_{p}$. Due to $\Omega _{RF}\sim |E_{RF}|$ and $F\left[ x \right] \sim T_p$, the normalized AM-AM transfer function $F\left[ x \right]$ can be expressed as
\begin{equation}
	\begin{aligned}
		F\left[ x \right] =e^{-\frac{ax^2}{b+x^2}},
	\end{aligned}
\end{equation}
where $a$ and $b$ are coefficients fitted based on experimental results, exhibiting numerically distinct values from their theoretical counterparts $\tilde{a}$ and $\tilde{b}$ derived from the simplified Doppler-free model. This AM-AM function describes the relationship between the RF intensity $x$ and the corresponding APD output voltage $F\left[ x \right]$.

\begin{figure}[htbp]
	\centering
	\includegraphics[width=0.45\textwidth]{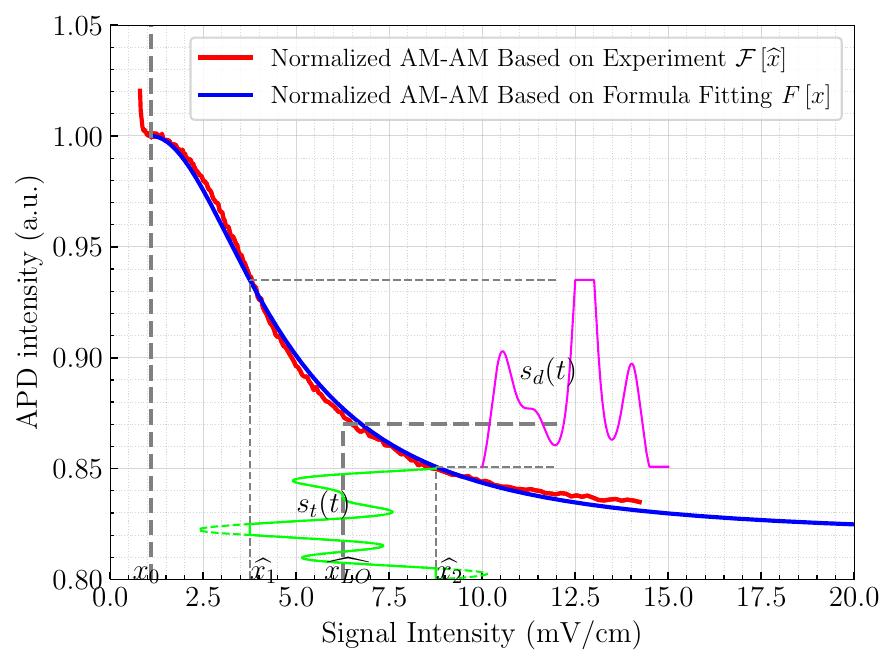}
	\caption{Red line: Normalized AM-AM based on experiment. Blue line: Normalized AM-AM based on formula fitting.}
	\label{AMAM}
\end{figure}

In Fig. \ref{AMAM}, we compare the normalized AM-AM curve based on experiment $\mathcal{F}\left[ \widehat{x} \right]$ and formula fitting $F\left[ x \right]$. Experimental measurements reveal a detection threshold $x_0$ in the Rydberg system, where microwave signals with intensity below $x_0$ fail to produce measurable responses. Therefore, the signals need to avoid working in this area. In addition, the blue line, fitted according to the data lager than $|E_{RF}|>x_0$, corresponds well to the experimental results, i.e., $\mathcal{F}\left[ x+x_0 \right] =F\left[ x \right] $ and $\widehat{x}=x+x_0$.

Considering the linear relationship between the experimentally measured AM-AM function and its normalized counterpart, without loss of generality, we mainly focus on normalized $F\left[ x \right]$ in the remainder of this paper.

\subsection{The Properties of $F[x]$}\label{T24}

We present three fundamental characteristics of the normalized AM-AM function $F[x]$:

\begin{itemize}
	\item[$\bullet$] The operating point of maximum slope
\end{itemize}
\begin{itemize}
	\item[$\bullet$] The convergence domain of Taylor expansion
\end{itemize}
\begin{itemize}
	\item[$\bullet$] The recursive property of Taylor expansion
\end{itemize}
where the maximum slope defines the optimal linear response, while the convergence domain and the recursive property enable quantitative analysis of OFDM nonlinearity in Sec. \ref{T3}.

\subsubsection{Maximum slope $|F^{(1)}[x]|$}\label{T231}

The slope $|F^{(1)}[x]|$, i.e., the first derivative of $F[x]$, is related to the signal gain. Consider $F^{(1)}[x]<0$ for $x>0$, the $x_{max}$ is given by

\begin{equation}
	\begin{aligned}
x_{max}&=\mathop{\arg\max}\limits_{x} |F^{\left( 1 \right)}\left[ x \right] |
\\
&=\mathop{\arg\min}\limits_{x} F^{\left( 1 \right)}\left[ x \right] ,
	\end{aligned}
\end{equation}
where $x_{max}^2=\frac{-b-ab+b\sqrt{a^2+2a+4}}{3}$. And for lager $a$, we have

\begin{equation}
	\begin{aligned}
x_{max}^2&=\frac{\sqrt{\left( a+1 \right) ^2+3}-\left( 1+a \right)}{3}b
\\
&=\frac{b\left( a+1 \right)}{3}\left( \sqrt{1+\frac{3}{\left( a+1 \right) ^2}}-1 \right) 
\\
&= \frac{b}{2\left( a+1 \right)}+\mathcal{O}\left( \frac{1}{\left( a+1 \right) ^2} \right) .
	\end{aligned}
\end{equation}

\subsubsection{The convergence domain of Taylor expansion} \label{T232}

For the given function $F\left[ x \right] =e^{-\frac{ax^2}{b+x^2}}$ with $a,b>0$, the Taylor expansion of $F\left[ x \right]$ in $x=x_{LO}$ is given by

\begin{equation}\label{Tayloreq}
	\begin{aligned}
		F\left[ x \right] =\sum_{m=0}^{+\infty}{\frac{F^{\left( m \right)}\left[ x_{LO} \right]}{m!}\left( x-x_{LO} \right) ^m},
	\end{aligned}
\end{equation}
where $F^{(m)}[x_{LO}]$ denote the $m$-th derivative of the $F[x]$ at $x=x_{LO}$. And $F\left[ x \right]$ can be expressed as the sum of series for any $x\in \mathbb{R}$ and $x\in (x_{LO}-\sqrt{x_{LO}^2+b},x_{LO}+\sqrt{x_{LO}^2+b})$ as $x_{LO}>0$ due to there are two singular point on the complex domain as $x=\pm \sqrt{-b}j$.

\subsubsection{The recursive property of Taylor expansion}\label{T233}

Denote $c_{m}[x_{LO}]=\frac{F^{\left( m \right)}\left[ x_{LO} \right]}{m!}$, the recursive formula is given by

\begin{equation}\label{theoremeq1}
	\begin{aligned}
		&mc_{m}[x_{LO}]+4x_{LO}\left( m+1 \right) c_{m+1}[x_{LO}]
		\\
		&+2\left( \left( m+2 \right) \left( 3x_{LO}^2+b \right) +ab \right) c_{m+2}[x_{LO}]
		\\
		&+2x_{LO}\left( 2\left( b+x_{LO}^2 \right) \left( m+3 \right) +ab \right) c_{m+3}[x_{LO}]
		\\
		&+\left( m+4 \right) \left( b+x_{LO}^2 \right) ^2c_{m+4}[x_{LO}]=0,
	\end{aligned}
\end{equation}
where the recurrence formula enables efficient Taylor expansion series about arbitrary operating points without requiring explicit computation of higher-order derivatives, significantly simplifying nonlinear system analysis. The complete mathematical derivation is presented in Appendix \ref{A1}.

\begin{figure}[htbp]
	\centering
	\includegraphics[width=0.45\textwidth]{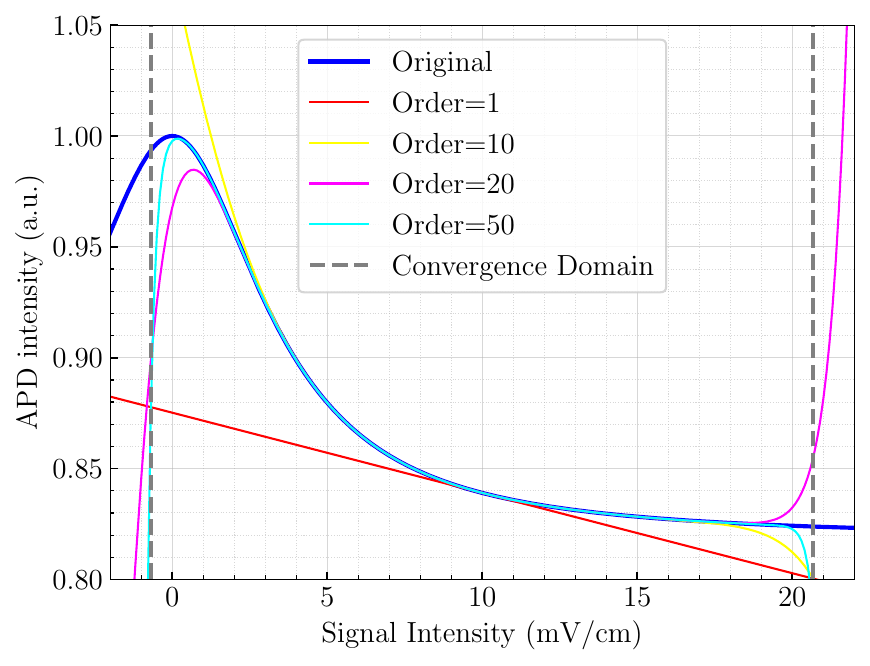}
	\caption{Taylor's convergence domain and higher order approximation.}
	\label{AMAM2}
\end{figure}

In Fig. \ref{AMAM2}, we display the convergence domain and higher order approximation of AM-AM function $F[x]$. The Taylor series expansions demonstrate convergence within this domain, with approximation accuracy systematically improving as the expansion order increases.

\section{Communication System Model}\label{T3}

\begin{figure*}
	\centering
	\includegraphics[width=0.95\textwidth]{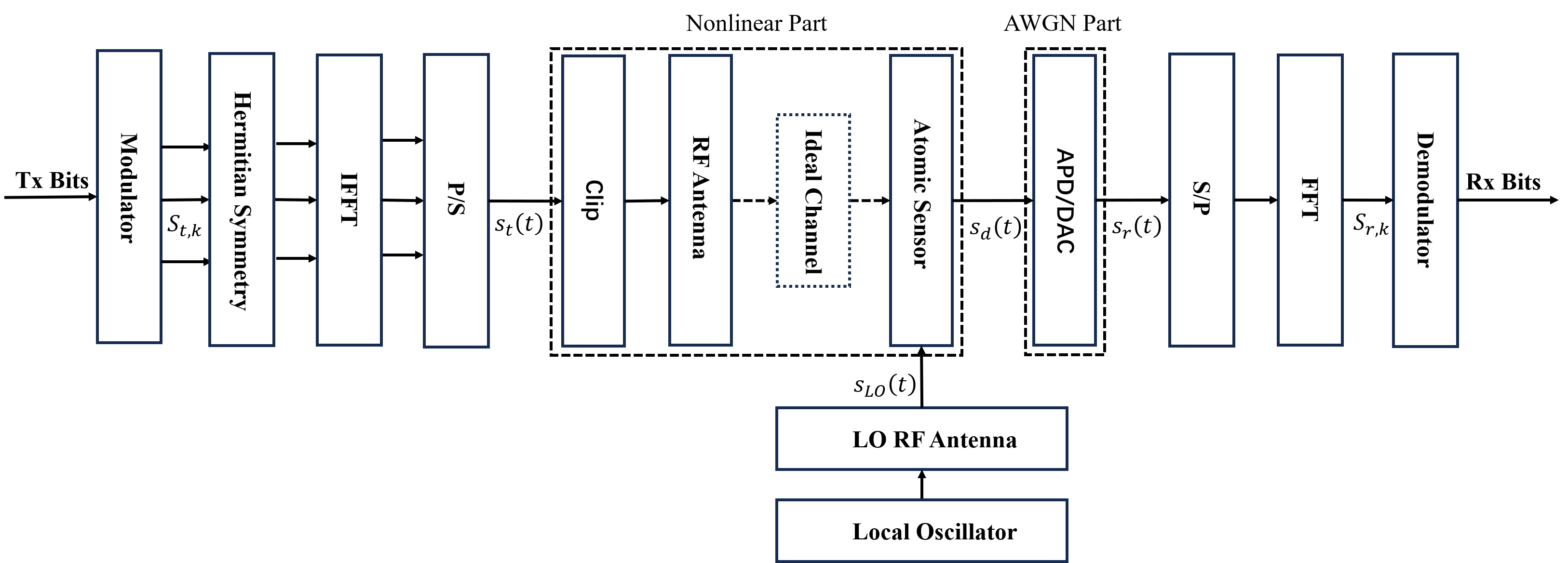}
	\caption{Block diagram of LODC-OFDM system.}
	\label{blockdiagram}
\end{figure*}

\begin{figure}[htbp]
	\centering
\includegraphics[width=0.45\textwidth]{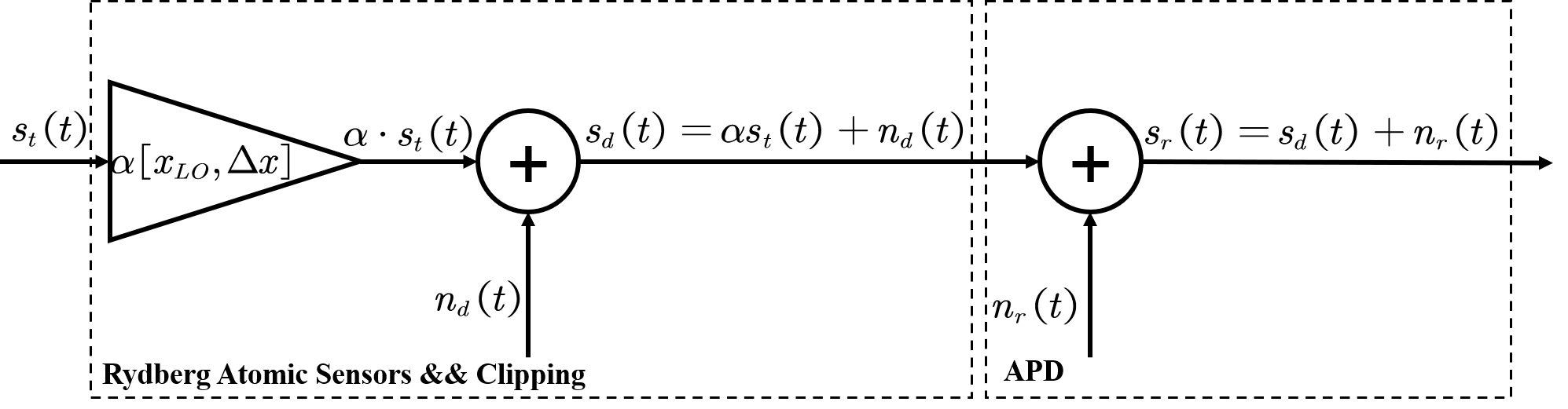}
\caption{Equivalent linear model for LODC-OFDM signal.}
\label{Equivalentmodel}
\end{figure}

\subsection{Introduction of the LODC-OFDM System}\label{T31}

The proposed LODC-OFDM (Local Oscillation Direct Current-biased Orthogonal Frequency Division Multiplexing) scheme builds upon conventional DCO-OFDM architecture. As illustrated in Fig. \ref{blockdiagram}, the system employs 2$N$ subcarriers with adaptive modulation parameters, leveraging the envelope detection characteristics of Rydberg sensors under low-bandwidth conditions. Under these limitations, the system's nonlinear distortion can be fully characterized by the AM-AM function $F\left[ x \right] $ derived in Sec. \ref{T23}.

At the transmitter, the bitstream is modulated into complex-valued symbols $S_{t,k}$ at subcarrier $k$. The original time-domain transmitted signal $s_{t}(t)$ is then obtained via the inverse fast Fourier transform (IFFT) of $S_{t,k}$. Denote $f_s=1/\Delta t$ be the symbol rate of information signal and $2N$ the number of carriers. The symbol period is $T_s=2N\Delta t=2N/f_s$ and the envelope of the LODC-OFDM signal $s_{t}(t)$ in each symbol can be expressed as 
\begin{equation}
	\begin{aligned}
s_t\left( t \right) =q_0\sum_{k=0}^{2N-1}{S_{t,k}e^{j2\pi f_kt}},
	\end{aligned}
\end{equation}
where $f_k=k\Delta f=kf_s/N$ and $S_{t,k}$ are the frequency and modulated symbol of the $k$-th carrier. $q_0$ is a scaling factor common to all carriers, we consider $q_0=1$ without losing generality.

Similar to the DCO-OFDM architecture, in order to obtain the real valued time-domain signal, Hermitian symmetry, i.e., $S_{t,k}=S_{t,N-k}^*$, is adopted for the OFDM frames, and $S_{t,0}$ and $S_{t,N}$ are set to zero. Furthermore, the LODC-OFDM scheme employs a local oscillator signal $s_{LO}(t)$ proximal to the Rydberg receiver, replacing the conventional DC bias used in optical DCO-OFDM transmitters, while maintaining equivalent functionality.

Denote the strength of local oscillator signal $s_{LO}(t)$ as $|s_{LO}(t)|=x_0+x_{LO}$. We adopt the double-sided clipping process before sending $s_{t}(t)$, and the equivalent received signal $s_{atom}\left( t \right)$ of the Rydberg sensor is given by,
\begin{equation}
	\begin{aligned}
	s_{atom}\left( t \right) &=\text{Clip}\left[ s_t\left( t \right) ,\Delta x \right] +s_{LO}\left( t \right) 
	\\
	&=
	\begin{cases}
		\Delta x+x_0+x_{LO} \quad \ \ s_t\left( t \right) >\Delta x\\
		s_t\left( t \right) +x_0+x_{LO} \quad -\Delta x\leq s_t\left( t \right) \leq \Delta x\\
		-\Delta x+x_0+x_{LO} \quad s_t\left( t \right) <-\Delta x
	\end{cases},
	\end{aligned}
\end{equation}
where the clipping operator $\text{Clip}\left[ s_t\left( t \right) ,\Delta x \right]$ characterizes the nonlinear signal processing applied to LODC-OFDM transmissions, producing equivalent thresholding effects at the receiver such that received envelope amplitudes are truncated to $\Delta x$ when exceeding an upper threshold $\Delta x$ or to $-\Delta x$ when falling below a lower threshold $-\Delta x$.

Signals exceeding the power threshold ($|s_t(t)|>\Delta x$) are clipped to ensure that when the local oscillator field satisfies $s_{t}(t)+|s_{LO}(t)|>x_0$ (i.e., $x_{LO} > \Delta x$), the received non-negative envelope signal $s_{atom}\left(t \right) $ consistently operates within the work area shown in Fig. \ref{AMAM}. Furthermore, under this configuration, according to the property of convergence domain in Sec. \ref{T232}, the signal envelope after clipping process satisfies
\begin{equation}
	\begin{aligned}
		x_{LO}-\sqrt{x_{LO}^2+b}<x_1<x_{LO}<x_2<x_{LO}+\sqrt{x_{LO}^2+b},
	\end{aligned}
\end{equation}
where $x_1=\widehat{x_1}-x_0=\widehat{x_{LO}}-x_0-\Delta x=x_{LO}-\Delta x$ and $x_2=\widehat{x_2}-x_0=\widehat{x_{LO}}-x_0+\Delta x=x_{LO}+\Delta x$. Thus, this clipping configuration ensures that the received signal, after superposition with the local oscillator electric field exceeding the original signal intensity, remains strictly within the convergence domain of the AM-AM function.

In summary, the LODC-OFDM architecture retains two fundamental features from conventional DCO-OFDM: Hermitian symmetry for real-valued signal generation and transmitter-side clipping. The key innovation lies in replacing the DC bias with a local oscillating electric field $x_0+x_{LO}$. These configurations guarantee that when the LO field strength exceeds the received signal intensity $x_{LO}>\Delta x$, the composite signal always remains within both the operational regime of the Rydberg sensor and the convergence domain of the AM-AM function's Taylor expansion.

From an alternative perspective, our selection of the DCO-OFDM architecture over ACO-OFDM for Rydberg sensor communications warrants careful consideration, despite the theoretical feasibility of adapting either approach. While ACO-OFDM offers the inherent advantage of DC bias elimination, the unique operational paradigm of Rydberg sensors fundamentally alters this trade-off analysis. Unlike conventional optical communications where DC bias originates from transmitter-side LED, Rydberg systems benefit from local oscillator electric fields generated near the sensor. This proximity advantage enables efficient DC bias provisioning without suffering from wireless channel attenuation, allowing substantial bias levels to be achieved with relatively low power consumption. More critically, the spectral efficiency penalty inherent to ACO-OFDM - specifically its 50$\%$ bandwidth utilization - proves particularly detrimental for Rydberg systems where instantaneous bandwidth represents a severely constrained resource. These combined factors establish DCO-OFDM as the superior architectural choice, primarily due to its full spectral utilization capability that better accommodates the stringent bandwidth requirements of Rydberg sensing applications.

\subsection{Nonlinearity Analysis of the LODC-OFDM System}\label{T32}

The LODC-OFDM system architecture, illustrated in Fig. \ref{blockdiagram}, operates under ideal channel conditions with two primary noise components: additive white Gaussian noise (AWGN) predominantly generated at the receiver, and nonlinear distortion arising from both the signal clipping process at the transmitter and the amplitude-dependent nonlinearity characterized by the AM-AM function $F\left[ x\right] $ at the receiver. Denote $s_d(t)$ be the signal after Rydberg sensor and clipping process, which is given by

\begin{equation}
	\begin{aligned}
s_d\left( t \right) &=\mathcal{F}\left[ s_{atom}\left( t \right) \right] 
\\
&=G\left[ s_t\left( t \right) \right] ,
	\end{aligned}
\end{equation}
where we establish a composite AM-AM nonlinear transfer function $G\left[ s_t\left( t \right) \right]$ that jointly characterizes both the transmitter-side clipping effects and receiver-side AM-AM distortion. The combined action of the local oscillator electric field and signal clipping maintains the composite signal $s_{atom}\left( t \right)$ within the Rydberg receiver's work area. Consequently, the overall system response to an original input signal $s_t\left(t \right) $ can be mathematically modeled as an uncut signal processed through an equivalent composite AM-AM transfer function $G\left[ s_t\left( t \right) \right]$, which is given by
\begin{equation}
	\begin{aligned}
&G\left[ x \right] =
\\
&\underset{\text{AM}-\text{AM\ Distortion}}{\underbrace{F\left[ x+x_{LO} \right] \left( u\left[ x-\left( x_1-x_{LO} \right) \right] -u\left[ x-\left( x_2-x_{LO} \right) \right] \right) }}+
\\
&\underset{\text{Clipping\ Distortion}}{\underbrace{F\left[ x_2 \right] u\left[ x-\left( x_2-x_{LO} \right) \right] +F\left[ x_1 \right] u\left[ -x+\left( x_1-x_{LO} \right) \right] }},
	\end{aligned}
\end{equation}
where $u\left[\cdot \right]$ is unit step function and $x_{LO}-x_1=x_{2}-x_{LO}=\Delta x$. The overall nonlinear response $G\left[ x \right]$ systematically incorporates distinct contributions from both transmitter and receiver subsystems. The nonlinearity at the receiver is reflected in the product of the difference between two step functions and AM-AM function $F\left[ x \right]$. While the clipping-induced nonlinearity at the transmitter is mathematically represented in the transfer function $G\left[ x \right]$ through two unit step functions. 

Figure \ref{Equivalentmodel} presents the equivalent linear representation of both the nonlinear components and additive white Gaussian noise depicted in Fig. \ref{blockdiagram}. The distorted signal $s_d(t)$ of nonlinear devices is modeled by making use of the Bussgang theorem \cite{Banelli2000Theoretical}, which gives the separateness of the nonlinear output as the sum of a useful attenuated input replica and an uncorrelated nonlinear distortion. Thus, $s_d(t)$ can be expressed as

\begin{equation}
	\begin{aligned}
s_d\left( t \right) &=s_u\left( t \right) +n_d\left( t \right) 
\\
&=\alpha \cdot s_t\left( t \right) +n_d\left( t \right), 
	\end{aligned}
\end{equation}
where $s_u\left( t \right)$ and $n_d\left( t \right)$ are the useful and distorted part of the signal $s_d\left( t \right)$. $\alpha = \alpha [x_{LO},\Delta x]$, denoted as attenuation factor, is related to the LO working point $x_{LO}$ and $\Delta x$.

The final received signal $s_r\left( t \right)$ is the sum of distorted signal $s_d(t)$ and uncorrelated additive white Gaussian noise $n_r(t)$, which is given by

\begin{equation}
	\begin{aligned}
s_r\left( t \right) &=s_d\left( t \right) +n_r\left( t \right) 
\\
&=\alpha \cdot  s_{t}\left( t \right) +n_d\left( t \right) +n_r\left( t \right) .
	\end{aligned}
\end{equation}

The transmitted signal $s_{t}(t)$ in each symbol is completely specified by $2N$ signal symbols $s_{t}=s_{t,n}\triangleq s_t(n\Delta t)$, which is given by 

\begin{equation}
	\begin{aligned}
		s_{t,n}=\sum_{k=0}^{2N-1}{S_{t,k}e^{j\frac{2\pi}{2N}nk}},
	\end{aligned}
\end{equation}
where $s_{t}$ can be approximated by a zero-mean Gaussian distribution with variance $\sigma_{t}^2$, when 2$N$ is larger than 64 according to the central limit theorem\cite{2000A}.

The conventional demodulator implements a $2N$-point Fast Fourier Transform (FFT) on the uniformly sampled received signal $s_{r}=s_{r,n} \triangleq s_{r}(n\Delta t)$, where $\Delta t$ represents the sampling interval. This spectral decomposition yields the decision variables at the demodulator input, which can be expressed as

\begin{equation}
	\begin{aligned}
		S_{r,k}=\frac{1}{2N}\sum_{n=0}^{2N-1}{s_{r,n}e^{-j\frac{2\pi}{2N}nk}},
	\end{aligned}
\end{equation}
where the received frequency-domain signal components $S_{r,k}$ would perfectly match their transmitted counterparts $S_{t,k}$ in the absence of nonlinear distortion and noise.

When accounting for the nonlinear distortion inherent to the proposed LODC-OFDM architecture, the time-domain distorted signal is sampled as $s_d=s_{d,n} \triangleq s_{d}(n\Delta t)$ and its frequency-domain representation obtained via $2N$-point FFT follows the modified Bussgang decomposition $S_{d,k}=\alpha \cdot S_{t,k}+N_{d,k}$. The final demodulated symbol $S_{r,k}$ at the $k$-th subcarrier can be fundamentally decomposed into the sum of three independent components, which are given by

\begin{equation}\label{eq1}
	\begin{aligned}
S_{r,k}&=S_{d,k}+N_{r,k}
\\
&=\alpha \cdot S_{t,k}+N_{d,k}+N_{r,k},
	\end{aligned}
\end{equation}
where $\alpha \cdot S_{t,k}$ in Eq. (\ref{eq1}) is the useful and undistorted data as the attenuation factor $\alpha$ is opportunely compensated in the end. The $N_{d,k}$ and $N_{r,k}$ terms, calculated by FFT of the distortion noise $n_d(t)$ and of the APD noise $n_r(t)$, are two uncorrelated noise and finally distort the received symbols in the receiver.

\subsubsection{Attenuation Factor $\alpha[x_{LO},\Delta x]$}\label{T321}

According to Bussghang theorem, the attenuation factor  of the useful component $\alpha [x_{LO},\Delta x]$ is given by
\begin{equation}
	\begin{aligned}
\alpha &=\frac{R_{s_ds_t}\left( 0 \right)}{R_{s_ts_t}\left( 0 \right)}
\\
&=\frac{\mathbb{E}\left\{ s_ds_t \right\}}{\mathbb{E}\left\{ s_ts_t \right\}}
\\
&=\frac{\mathbb{E}\left\{ G\left[ s_t \right] s_t \right\}}{\mathbb{E}\left\{ s_ts_t \right\}},
	\end{aligned}
\end{equation}
where $R_{s_ds_t}(\tau )$ denotes the cross-correlation function between the distorted signal $s_d$ and original transmitted signal $s_t$. $R_{s_ts_t}(\tau )$ is the auto-correlation function of transmitted signal $s_t$, where the zero-lag value  $R_{s_ts_t}(0 )=\mathbb{E}\left\{ s_ts_t \right\}=\sigma_t^2$ corresponds to the total signal power. And $\mathbb{E}\left\{ \cdot \right\} $ denotes the statistical expectation operator.

Based on the rapid $\left( 2I+1\right) $-th order Taylor expansion of the AM-AM function provided in Eq. (\ref{Tayloreq}) and Eq. (\ref{theoremeq1}), the attenuation factor $\alpha$ can be approximated as the summation of $I+1$ dominant terms, expressed as

\begin{equation}\label{alphaeq}
	\begin{aligned}
\alpha =\sum_{i=0}^I{\left( 2i+1 \right) c_{2i+1}\left[ x_{LO} \right] v_{2i}},
	\end{aligned}
\end{equation}
where $v_{2i}$ is given by
\begin{equation}
	\begin{aligned}
		v_{2i}=\sigma _{t}^{2i}\left( 2i-1 \right) !!-\frac{2^i\sigma _{t}^{2i}}{\sqrt{\pi}}W\left( i+\frac{1}{2},\widetilde{\Delta x}^2 \right) ,
	\end{aligned}
\end{equation}
where $(2i-1)!!$ denotes double factorial and $(-1)!!=1$. $W\left( s,x \right) =\int_x^{+\infty}{t^{s-1}e^tdt}$ is the incomplete gamma function and $\widetilde{\Delta x}=\frac{\Delta x}{\sqrt{2}\sigma_{t}}$. The complete mathematical derivation of these results is presented in Appendix \ref{A2}.

\subsubsection{Nonlinear and Clipping Noise $\sigma_{d}^2$}\label{T322}

The nonlinear AM-AM function $F[x]$ and clipping process not only reduces the useful signal power by affecting attenuation factor $\alpha$, but also introduce noise $n_d$ with power $\sigma_{d}^2$. Given that the transmitted signal $s_t$ follows a zero-mean Gaussian distribution $\mathbb{E}\left\{ s_t \right\}=0$, the first-order statistic of the nonlinear distortion noise $n_d = s_d - \alpha s_t$ simplifies to $\mathbb{E}\left\{ n_d \right\}=\mathbb{E}\left\{ s_d-\alpha s_t \right\}=\mathbb{E}\left\{ s_d \right\}$, as the linear term $\alpha \mathbb{E}\left\{s_t\right\}$ vanishes. Consequently, the nonlinear noise power $\sigma _{d}^{2}$ is given by

\begin{equation}
	\begin{aligned}
		\sigma _{d}^{2}&=\mathbb{D}\left\{n_d\right\}
		\\
		&=\mathbb{E}\left\{ n_{d}^{2} \right\} -\mathbb{E}\left\{ n_d \right\} ^2
		\\
		&=\mathbb{E}\left\{ s_{d}^{2} \right\} -\alpha ^2\mathbb{E}\left\{ s_{t}^{2} \right\} -\mathbb{E}\left\{ n_d \right\} ^2
		\\
		&=\mathbb{E}\left\{ s_{d}^{2} \right\}  -\mathbb{E}\left\{ n_d \right\} ^2-\alpha ^2\sigma_{t}^2,
	\end{aligned}
\end{equation}
where the attenuation factor $\alpha$ can be computed through the closed-form expression given in Eq. (\ref{alphaeq}) and signal power $\sigma_{t}^2$ is known parameter in our analysis. Based on the rapid $\left( 2I\right) $-th order Taylor expansion provided in Eq. (\ref{theoremeq1}), $\mathbb{E}\left\{ s_{d}^{2} \right\}$ and $\mathbb{E}\left\{ n_d \right\}$ can also be approximated as the summation of $I+1$ dominant terms, which are given by
\begin{equation}\label{noieseq}
	\begin{aligned}
\mathbb{E}\left\{ n_d \right\} &=\sum_{i=0}^I{c_{2i}\left[ x_{LO} \right] v_{2i}}
\\
&\ \ +\left( F\left[ x_{LO}-\Delta x \right] +F\left[ x_{LO}+\Delta x \right] \right) Q\left( \sqrt{2}\widetilde{\Delta x} \right) ,
\\
\mathbb{E}\left\{ s_{d}^{2} \right\} &=\sum_{i=0}^I{\tilde{c}_{2i}\left[ x_{LO} \right] v_{2i}}
\\
&\ \ +\left( F^2\left[ x_{LO}-\Delta x \right] +F^2\left[ x_{LO}+\Delta x \right] \right) Q\left( \sqrt{2}\widetilde{\Delta x} \right) ,
	\end{aligned}
\end{equation}
where $\tilde{c}_{2i}\left[ x_{LO} \right] =\sum_{k=0}^{2i}{c_k\left[ x_{LO} \right] c_{2i-k}\left[ x_{LO} \right]}$ and $Q\left( x \right) =\int_x^{+\infty}{\frac{1}{\sqrt{2\pi}}e^{-t^2/2}dt}$. The complete mathematical derivation of these results is presented in Appendix \ref{A3}.

Therefore, based on the approximation of the $I+1$ terms, attenuation factor $\alpha$, $\mathbb{E}\left\{ s_{d}^{2} \right\}$ and $\mathbb{E}\left\{ n_d \right\}$ can be approximated, and finally noise power $\sigma _{d}^{2}$ can be calculated.

\subsection{Performance Analysis of the LODC-OFDM System}\label{T33}

The computational framework developed in Sec. \ref{T32} enables efficient evaluation of the critical parameters including attenuation factor $\alpha$ and nonlinear noise power $\sigma _{d}^{2}$. This allows comprehensive characterization of the proposed LODC-OFDM architecture through two fundamental performance including signal-to-noise ratio (SNR), and BER.

The nonlinear distortion term $N_{d,k}$, obtained via FFT processing, comprises a superposition of numerous independent terms and, by the Central Limit Theorem, converges to a zero-mean complex Gaussian distribution with variance $\sigma_{d,k}^2 = \frac{\sigma_d^2}{2N}$, where the approximation accuracy improves with increasing subcarrier count $N$. Analogously, the APD noise follows $\sigma_{r,k}^2 = \frac{\sigma_r^2}{2N}$ per subcarrier. Under $M$-QAM modulation, the transmitted symbols $S_{t,k}$ constitute independent and identically distributed random variables drawn uniformly from $M$-ary constellations, with per-subcarrier power $\sigma_{t,k}^2 = \frac{2(M-1)}{3}$. Consequently, the aggregate transmit power across all $2N-2$ active subcarriers is $\sigma_t^2 =\frac{2q_0^2(M-1)(2N-2)}{3}$.

\subsubsection{SNR}
The SNR degradation induced by nonlinear distortion and clipping effects is quantified by $SNR_d$, corresponding to the SNR of the distorted signal $s_d\left( t\right) $ in Fig. \ref{blockdiagram}, while $SNR_r$ represents the total system SNR incorporating both nonlinear effects and receiver-added Gaussian noise, as measured from $s_r\left( t\right)$ in Fig. \ref{blockdiagram}. Extending this framework to individual subcarriers, we define $SNR_{d,k}$ and $SNR_{r,k}$ as the per-subcarrier SNRs for $s_d\left( t\right)$ and $s_r\left( t\right)$, respectively. We have

\begin{equation}
	\begin{aligned}
		SNR_{d,k}&=\frac{\alpha ^2\sigma _{t,k}^{2}}{\sigma _{d,k}^{2}}
		\\
		&=\frac{N}{ N-1 }\cdot \frac{\alpha ^2\sigma _{t}^{2}}{\sigma _{d}^{2}}
		\\
		&=\frac{N}{ N-1 }\cdot SNR_d
		\\
		SNR_{r,k}&=\frac{\alpha ^2\sigma _{t,k}^{2}}{\sigma _{d,k}^{2}+\sigma _{r,k}^{2}}
		\\
		&=\frac{N}{N-1}\cdot \frac{\alpha ^2\sigma _{t}^{2}}{\sigma _{d}^{2}+\sigma _{r}^{2}}
		\\
		&=\frac{N}{ N-1 }\cdot SNR_r,
	\end{aligned}
\end{equation}
where $SNR_{d,k}\approx SNR_{d}$ and $SNR_{r,k}\approx SNR_{r}$ for lager $N$.

%

\subsubsection{BER Performance}

The BER for arbitrary modulation schemes can be derived from the system's SNR. For the $M$-QAM modulation adopted in this work, the $k$-th carriers' BER induced by nonlinear effects $BER_{d,k}$ and the overall system $BER_{r,k}$ can be mathematically expressed as follows,
\begin{equation}\label{BEReq}
	\begin{aligned}
		BER_{d,k}&=\frac{4}{\log _2M}\left( 1-\frac{1}{\sqrt{M}} \right) Q\left(q_0 \frac{\sqrt{2}|\alpha |}{\sigma _{d,k}} \right) 
		\\
		BER_{r,k}&=\frac{4}{\log _2M}\left( 1-\frac{1}{\sqrt{M}} \right) Q\left(q_0 \sqrt{\frac{2\alpha ^2}{\sigma _{d,k}^{2}+\sigma _{r,k}^{2}}} \right) .
	\end{aligned}
\end{equation}

\section{NUMERICAL AND EXPERIMENTAL RESULTS}\label{T4}

The numerical results presented in this study are derived from the experimental measurements about AM-AM function $F\left[x \right] $ shown in Fig. \ref{AMAM}, with parameter a maintained at a constant value of $a=3.5609$ and $b=55.1572$.

\subsection{Attenuation Factor $\alpha$ and Nonlinear and clipping noise $\sigma_{d}^2$}\label{T41}
\begin{figure}[htbp]
	\centering
	\includegraphics[width=0.45\textwidth]{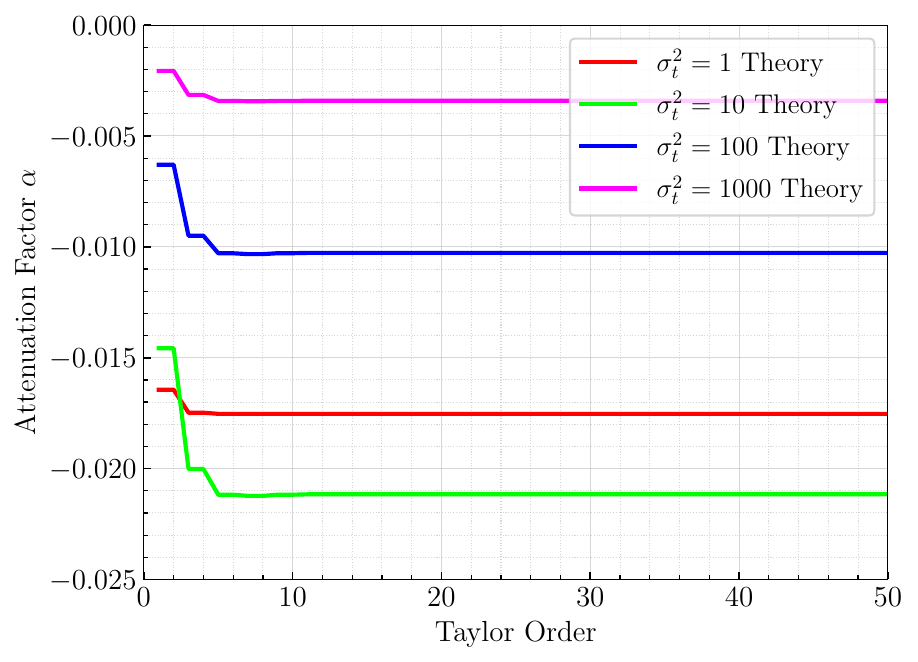}
	\caption{Attenuation Factor $\alpha$ with respective to Taylor order.}
	\label{Alpha_Order}
\end{figure}
\begin{figure}[htbp]
	\centering
	\includegraphics[width=0.45\textwidth]{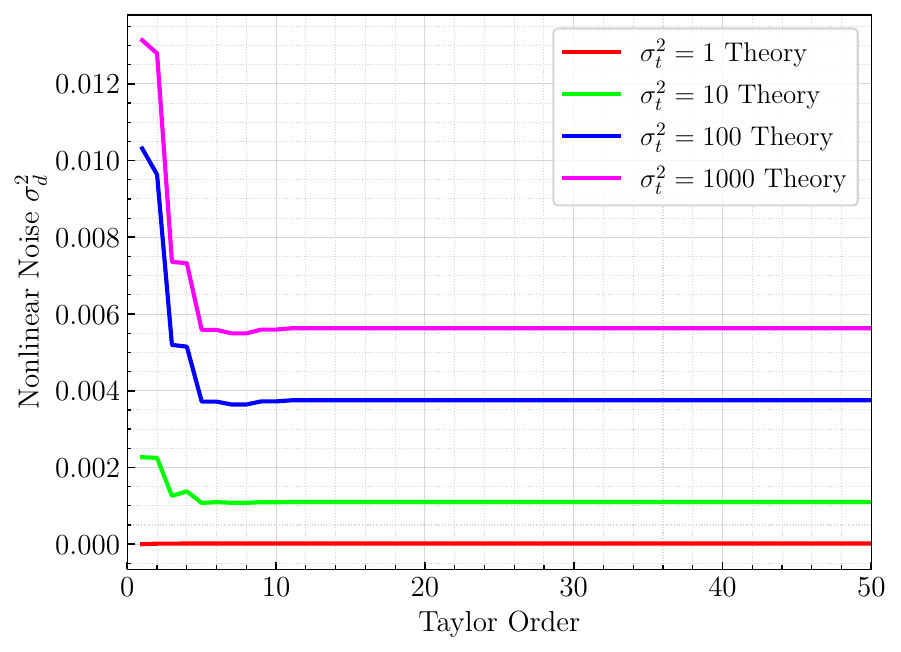}
	\caption{Nonlinear and clipping noise $\sigma_{d}^2$ with respective to Taylor order.}
	\label{ClipNoise_Order}
\end{figure}

In Fig. \ref{Alpha_Order} and Fig. \ref{ClipNoise_Order}, we present a systematic convergence analysis of both Eq. (\ref{alphaeq}) and Eq. (\ref{noieseq}) formulations under fixed local oscillator conditions $x_{LO}=10$ mV/cm and $\Delta x = 5$ mV/cm, demonstrating their respective convergence behaviors across four distinct signal power $\sigma_{t}^2$. The results quantitatively characterize the evolution of both the attenuation coefficient $\alpha$ and nonlinear noise $\sigma_{d}^2$ as functions of Taylor expansion order. Notably, the analysis reveals that convergence is effectively achieved for both formulations when the Taylor expansion order exceeds 10.

\subsection{SNR Performance}\label{T42}

\begin{figure}[htbp]
	\centering
	\includegraphics[width=0.45\textwidth]{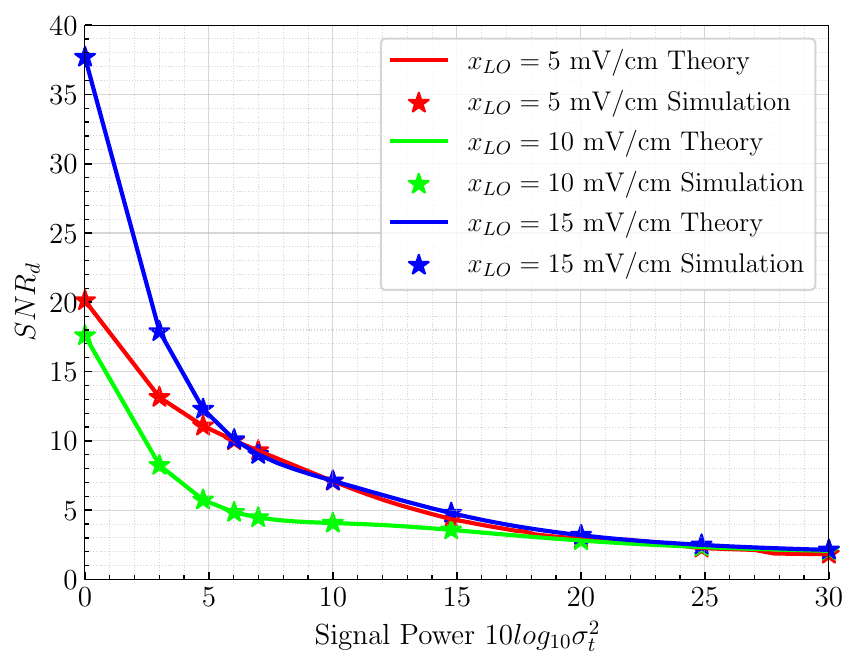}
	\caption{$SNR_{d}$ with respective to LODC-OFDM signal power $\sigma_{t}^2$.}
	\label{SNR_Cxx1}
\end{figure}

In Fig. \ref{SNR_Cxx1}, we systematically investigates the impact of input power $\sigma_{t}^2$ on $SNR_d$ degradation induced by nonlinear effects under ideal predistortion conditions $\sigma_{r}=0$. Three distinct local oscillator field strengths $x_{LO}=5$ mV/cm, $x_{LO}=10$ mV/cm, $x_{LO}=15$ mV/cm are examined, represented by red, green, and blue curves respectively, while maintaining $\Delta x =5$ mV/cm as a fixed parameter. The results reveal two distinct operational regimes. First, at lower input power $3\sigma_{t}<\Delta x$, i.e., $10log_{10}\sigma_{t}^2<4.437$, nonlinear noise $\sigma_{d}^2$ predominantly originates from higher-order nonlinearities in the AM-AM response function $F\left[ x_{LO}\right] $ rather than first-order components $F^{\left(1 \right)} \left[ x_{LO}\right] $. For instance, while $x_{LO}=5$ mV/vm exhibits steeper slope in the AM-AM response, its pronounced higher-order nonlinearities result in lower SNR compared to $x_{LO}=15$ mV/cm. Similarly, $x_{LO}=10$ mV/cm demonstrates the most severe SNR degradation due to significant curvature in its response characteristics as indicated in Fig. \ref{AMAM}. Secondly, in the high input power regime, i.e., $10log_{10}\sigma_{t}^2>4.437$, clipping-induced distortion becomes the dominant noise mechanism, leading to comparably poor SNR performance across all three $x_{LO}$ configurations.

\begin{figure}[htbp]
	\centering
	\includegraphics[width=0.45\textwidth]{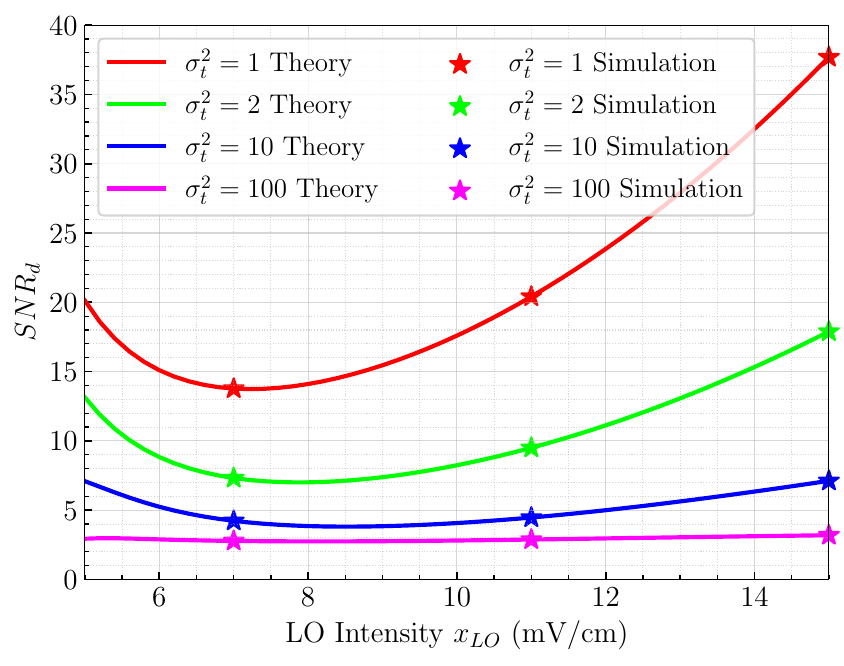}
	\caption{$SNR_{d}$ with respective to LO intensity $x_{LO}$.}
	\label{SNR_xLO1}
\end{figure}

In Fig. \ref{SNR_xLO1}, we further investigates the impact of LO intensity $x_{LO}$ on $SNR_d$ degradation induced by nonlinear effects. Consequently, when the back-end noise $\sigma_{r}^2=0$ becomes negligible, the system's SNR can be approximated as $SNR_d \approx SNR_r$. Under this condition, for low-power signals ($3\sigma_{t}<\Delta x$, represented by the red and green curves), the dominant factor affecting SNR is the AM-AM response function $F\left[x \right] $ rather than the clipping process. This suggests that optimizing the operating point of the local oscillator electric field can enhance SNR performance. Specifically, when $x_{LO}$ is below 7 mV/cm, the steeper slope of the AM-AM response function increases signal gain, thereby improving SNR. Conversely, for $x_{LO}$ values exceeding 11 mV/cm, although the AM-AM response exhibits a shallower slope (resulting in lower amplification gain), the function $F\left[x \right] $ becomes more linear, manifested as significantly reduced higher-order Taylor expansion components relative to the first-order term. This improved linearity minimizes nonlinear distortion noise $\sigma_{d}^2$, ultimately leading to superior overall SNR despite the reduced gain $\alpha$. For high-power signals ($3\sigma_{t}>\Delta x$, represented by the blue and purple curves), the dominant factor affecting SNR is clipping process, resulting the adjustment of the $x_{LO}$ operating point ineffective for significant SNR enhancement.

\begin{figure}[htbp]
	\centering
	\includegraphics[width=0.45\textwidth]{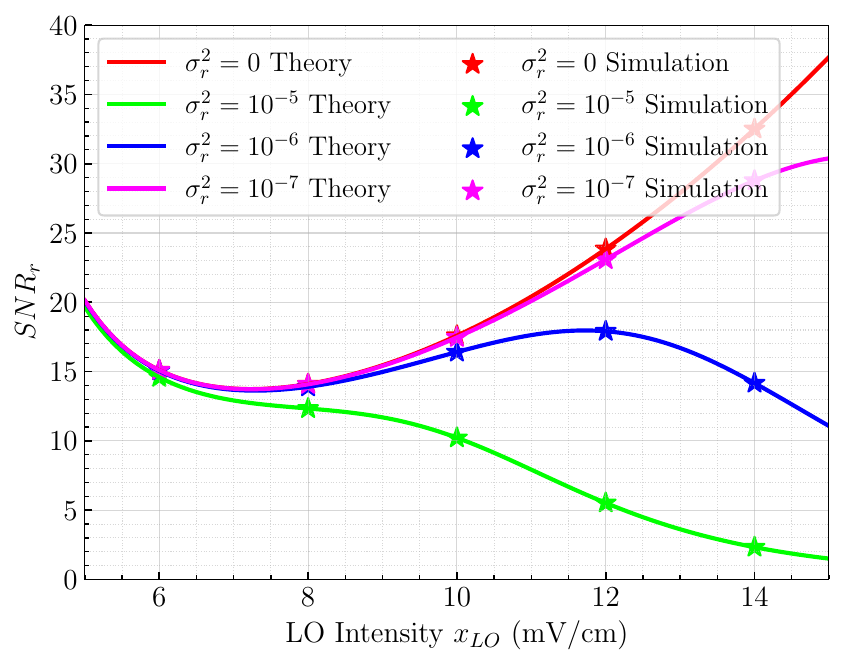}
	\caption{$SNR_{r}$ with respective to LO intensity $x_{LO}$.}
	\label{SNR_xLO2}
\end{figure}

However, when the back-end noise becomes non-negligible $\sigma_{r}^2 \neq0$, as demonstrated in Fig. \ref{SNR_xLO2}, a distinct trade-off emerges in system performance. For larger $x_{LO}$ values, while the system exhibits reduced nonlinear noise $s^2_d$, the diminished slope simultaneously results in a smaller attenuation factor $\alpha$, leading to decreased useful signal power $\alpha^2\sigma_{t}^2$ and consequently lower SNR. This analysis reveals an important optimization principle for the complete LODC-OFDM receiver system. When the back-end noise $\sigma_{r}$ can be negligible, operating at higher local oscillator field strengths yields superior SNR performance. But under conditions of significant back-end noise, optimal performance is achieved by biasing the local oscillator field at the region of maximum slope, thereby maximizing both the effective signal power $\alpha^2\sigma_{t}^2$ and overall $SNR_r$.

\subsection{BER Performance}\label{T43}

We configure the system parameters with $q_0 = 0.063$, employing an OFDM symbol length of $N = 128$ subcarriers, resulting in a total of $2N = 256$ carriers. The theoretical BER performance is derived from the analytical expression in Eq. (\ref{BEReq}), while the simulated results are obtained through Monte Carlo trials using randomly generated data sequences. Under this configuration, the signal power for three different QAM modulation orders $M=4, 16, 64$ are 2.0163, 10.0813, and 42.3413, respectively.

\begin{figure}[htbp]
	\centering
	\includegraphics[width=0.45\textwidth]{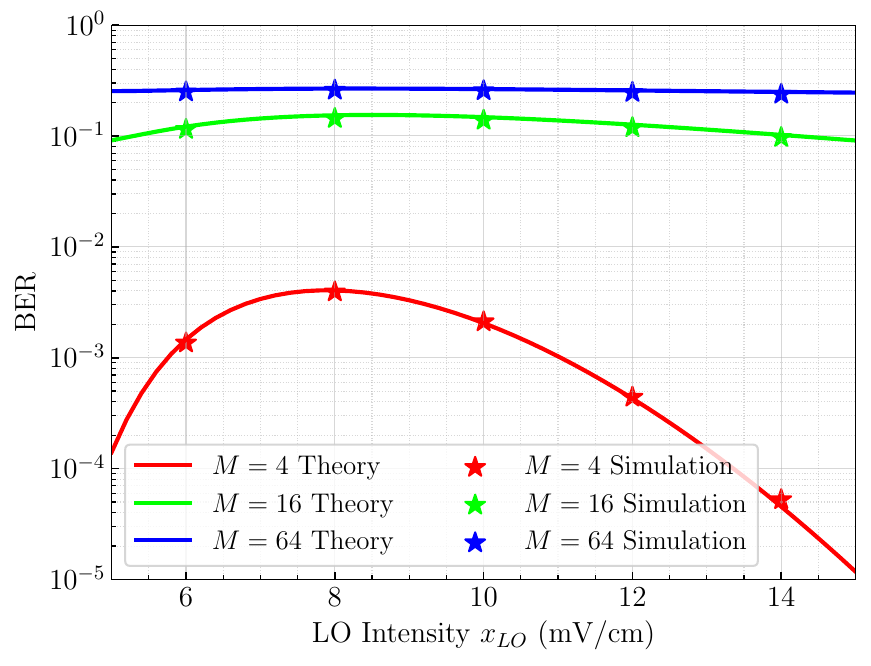}
	\caption{BER with respective to LO intensity $x_{LO}$.}
	\label{BER_xLO1}
\end{figure}

In Fig. \ref{BER_xLO1}, we presents the BER characteristics as a function of local oscillator field strength $x_{LO}$ for three modulation orders $M$, under the condition of negligible back-end noise $\sigma_{r}^2=0$. The results demonstrate distinct noise-dominated regimes. For the high-power signal $M=$ 16 and 64, clipping-induced distortion becomes the predominant noise mechanism, resulting in elevated BER levels that show minimal variation with LO field strength. Conversely, for the low-power signal, the BER exhibits strong dependence on LO field strength due to dominant nonlinearities originating from the AM-AM conversion characteristics.

\begin{figure}[htbp]
	\centering
	\includegraphics[width=0.45\textwidth]{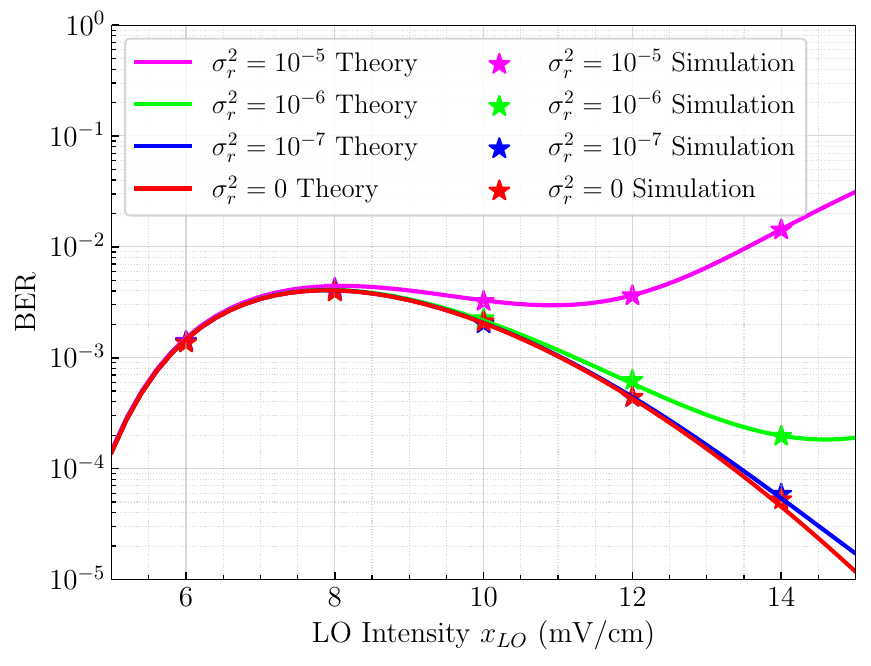}
	\caption{BER with respective to LO intensity $x_{LO}$.}
	\label{BER_xLO2}
\end{figure}

In Fig. \ref{BER_xLO2}, we analyzes theBER characteristics for low-power M=4 QAM signals as a function of both the local oscillator field strength $X_{lo}$ and back-end noise level $\sigma_{r}^2$. When $\sigma_{r}^2$ becomes non-negligible, excessive LO field strength leads to significant BER degradation. This occurs because while higher $x_{LO}$ values typically reduce nonlinear distortion, they simultaneously amplify the impact of back-end noise through decreased signal gain, ultimately compromising system performance.

\begin{figure*}[htbp]
	\centering
	\includegraphics[width=0.95\textwidth]{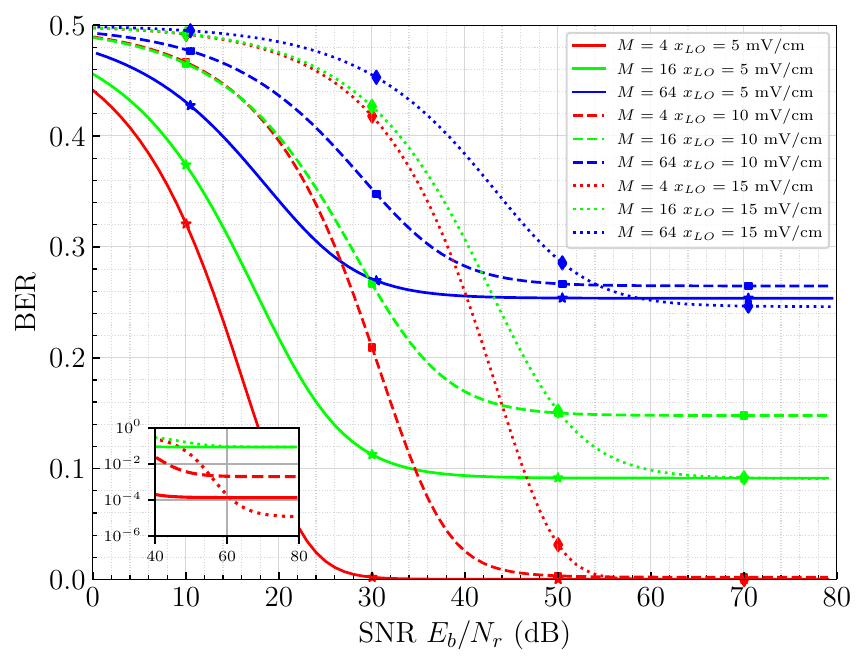}
	\caption{BER with respective to SNR $\frac{E_b}{N_r}$.}
	\label{PYBER_SNR1}
\end{figure*}

In Fig. \ref{PYBER_SNR1}, we presents the normalized bit energy $E_b$/$N_r$ characteristics for three distinct QAM modulation orders $M$, which is given by
\begin{equation}
	\begin{aligned}
		\frac{E_b}{N_r}=\frac{\sigma_{t}^2}{log_2M \times \sigma_{r}^2}.
	\end{aligned}
\end{equation}
where this normalization enables a systematic comparison of power efficiency across different modulation schemes under equivalent noise conditions. The curves of the same color represent the same modulation order, and the results indicate that for high $E_b$/$N_r$, a larger $x_{LO}$ will result in better BER performance, while for low $E_b$/$N_r$, a lower $x_{LO}$ is required. The reasons for this are explained in detail in Sec. \ref{T42}. The same linear curve represents the same local oscillator strength $x_{LO}$, and due to the small nonlinear distortion $\sigma_{d}^2$ of low-power signals, it has a lower BER.


\section{Conclusion} \label{T5}

In this work, we focus on the broadband OFDM reception based on Rydberg atomic receivers and its nonlinear performance analysis. We derived an approximation AM-AM distortion function based on the four-level Rydberg atoms physical model and detection system. Furthermore, we propose the LODC-OFDM scheme suitable for Rydberg sensing, which solves the problem of broadband OFDM signal reception. Then, we adopt Bussgang theorem to analyze the nonlinear distortion of LODC-OFDM signals and the results in closed-form solutions are derived for AM-AM curves approximated by Taylor series expansion and for the ideal pre-distortion case. In real experiments, the experimental and theoretical results fit well.

\section{Acknowledgement}
The authors would like to thank Mr. Guo-Yong Xiang, Mr. Kang-Da Wu, and Mr. Xie Chongwu for discussions and experimental guidance.

\appendices
\section{The recursive property of Taylor expansion}\label{A1}

For the given normalization AM-AM function $F[x]$ and its first derivative $F^{(1)}\left[ x \right]$, we have
\begin{equation}
	\begin{aligned}
		F\left[ x \right] &=e^{-\frac{ax^2}{b+x^2}},
\\
		F^{(1)}\left[ x \right] &=-\frac{2abx}{\left( b+x^2 \right) ^2}e^{-\frac{ax^2}{b+x^2}}.
	\end{aligned}
\end{equation}

After some algebra, we have
\begin{equation}
	\begin{aligned}
		\left( b+x^2 \right) ^2F^{\left( 1 \right)}\left[ x \right] =-2abxF\left[ x \right] ,
	\end{aligned}
\end{equation}

Based on Leibniz formula with $\left( fg \right) ^{\left( m \right)}=\sum_{k=0}^m{C_{m}^{k}f^{\left( k \right)}g^{\left( m-k \right)}}$, we have 

\begin{equation}
	\begin{aligned}
		&C_{m}^{0}\left( b+x^2 \right) ^2F^{\left( m+1 \right)}\left[ x \right] 
		\\
		&+\left( C_{m}^{1}4x\left( b+x^2 \right) +C_{m}^{0}2abx \right) F^{\left( m \right)}\left[ x \right] 
		\\
		&+\left( C_{m}^{2}\left( 12x^2+4b \right) +C_{m}^{1}2ab \right) F^{\left( m-1 \right)}\left[ x \right] 
		\\
		&+C_{m}^{3}24xF^{\left( m-2 \right)}\left[ x \right] +C_{m}^{4}24F^{\left( m-3 \right)}\left[ x \right] =0.
	\end{aligned}
\end{equation}

After some algebra, we have
\begin{equation}
	\begin{aligned}
		&\frac{\left( m+3 \right) \left( m+2 \right) \left( m+1 \right) m}{4!}24F^{\left( m \right)}\left[ x \right] 
		\\
		&+\frac{\left( m+3 \right) \left( m+2 \right) \left( m+1 \right)}{3!}24xF^{\left( m+1 \right)}\left[ x \right] 
		\\
		&+\left( \frac{\left( m+3 \right) \left( m+2 \right)}{2!}\left( 12x^2+4b \right) +\frac{\left( m+3 \right)}{1!}2ab \right) F^{\left( m+2 \right)}\left[ x \right] 
		\\
		&+\left( \frac{\left( m+3 \right)}{1!}4x\left( b+x^2 \right) +2abx \right) F^{\left( m+3 \right)}\left[ x \right] 
		\\
		&+\left( b+x^2 \right) ^2F^{\left( m+4 \right)}\left[ x \right] =0.
	\end{aligned}
\end{equation}

After simplification, we have
\begin{equation}
	\begin{aligned}
		&m\frac{F^{\left( m \right)}\left[ x \right]}{m!}+4x\left( m+1 \right) \frac{F^{\left( m+1 \right)}\left[ x \right]}{\left( m+1 \right) !}
		\\
		&+2\left( \left( m+2 \right) \left( 3x^2+b \right) +ab \right) \frac{F^{\left( m+2 \right)}\left[ x \right]}{\left( m+2 \right) !}
		\\
		&+2x\left( 2\left( b+x^2 \right) \left( m+3 \right) +ab \right) \frac{F^{\left( m+3 \right)}\left[ x \right]}{\left( m+3 \right) !}
		\\
		&+\left( m+4 \right) \left( b+x^2 \right) ^2\frac{F^{\left( m+4 \right)}\left[ x \right]}{\left( m+4 \right) !}=0,
	\end{aligned}
\end{equation}
where we denote $c_m[x]=\frac{F^{\left( m \right)}\left[ x \right]}{m!}$, the Taylor expansion coefficient can be solved by the above recursive process shown in Eq. (\ref{theoremeq1}). And the first four items can be expressed as

\begin{equation}
	\begin{aligned}
c_0\left[ x \right] &=e^{-\frac{ax^2}{b+x^2}}
\\
c_1\left[ x \right] &=-\frac{2abx}{\left( b+x^2 \right) ^2}e^{-\frac{ax^2}{b+x^2}}
\\
c_2\left[ x \right] &=\frac{ab\left( 3x^4+2\left( a+1 \right) bx^2-b^2 \right)}{\left( b+x^2 \right) ^4}e^{-\frac{ax^2}{b+x^2}}
\\
c_3\left[ x \right] &=-\frac{2abx}{3\left( b+x^2 \right) ^6}\left( 2a^2b^2x^2-3ab^3+6ab^2x^2+ \right. 
\\
&\left. 9abx^4-6b^3-6b^2x^2+6bx^4+6x^6 \right) e^{-\frac{ax^2}{b+x^2}}.
	\end{aligned}
\end{equation}

\section{Attenuation Factor $\alpha[x_{LO},\Delta x]$}\label{A2}

Given equivalent gain $G[x]$ and its first derivative $G^{(1)}\left[ x \right]$, we have
\begin{equation}
	\begin{aligned}
&G^{\left( 1 \right)}\left[ x \right] =
\\
&F^{\left( 1 \right)}\left[ x+x_{LO} \right] \left( u\left[ x-\left( x_1-x_{LO} \right) \right] -u\left[ x-\left( x_2-x_{LO} \right) \right] \right) 
\\
&+F\left[ x+x_{LO} \right] \left( \delta \left[ x-\left( x_1-x_{LO} \right) \right] -\delta \left[ x-\left( x_2-x_{LO} \right) \right] \right) 
\\
&+F\left[ x_2 \right] \delta \left[ x-\left( x_2-x_{LO} \right) \right] -F\left[ x_1 \right] \delta \left[ -x+\left( x_1-x_{LO} \right) \right], 
	\end{aligned}
\end{equation}
where $\delta\left[  \cdot\right] $ is Dirac delta function.

According to central limit theorem , $x$, i.e., $s_t$, follows a zero mean Gaussian distribution with variance $R_{s_{t}s_{t}}=\mathbb{E}\left\{ s_{t}s_{t} \right\}=\sigma_{t}^2$ and its probability distribution function $p\left( x \right) =\frac{1}{\sqrt{2\pi \sigma _{t}^{2}}}e^{-\frac{x^2}{2\sigma _{t}^{2}}}$. Thus, we have
\begin{equation}
	\begin{aligned}
		\alpha &=\frac{\mathbb{E}\left\{ G\left[ s_t \right] s_t \right\}}{\mathbb{E}\left\{ s_ts_t \right\}}
		\\
		&=\mathbb{E}\left\{ G^{\left( 1 \right)}\left[ s_t \right] \right\} 
		\\
		&=\int_{-\infty}^{+\infty}{G^{\left( 1 \right)}\left[ x \right] p\left( x \right) dx}
		\\
		&=\int_{-\infty}^{+\infty}{F^{\left( 1 \right)}\left[ x+x_{LO} \right] u\left[ x-\left( x_1-x_{LO} \right) \right] p\left( x \right) dx}
		\\
		&\ \ \ -\int_{-\infty}^{+\infty}{F^{\left( 1 \right)}\left[ x+x_{LO} \right] -u\left[ x-\left( x_2-x_{LO} \right) \right] p\left( x \right) dx}
		\\
		&\ \ \ +F\left[ x_1 \right] p\left( x_1 \right) -F\left[ x_2 \right] p\left( x_2 \right)
		\\
		 &\ \ \ +F\left[ x_2 \right] p\left( x_2 \right) -F\left[ x_1 \right] p\left( x_1 \right) 
		\\
		&=\int_{x_1-x_{LO}}^{x_2-x_{LO}}{F^{\left( 1 \right)}\left[ x+x_{LO} \right] p\left( x \right) dx}
\\
		&=\frac{1}{\sqrt{2\pi}\sigma _{t}}\sum_{m=0}^{+\infty}{\left( n+1 \right) c_{m+1}\left[ x_{LO} \right] \int_{x_1-x_{LO}}^{x_2-x_{LO}}{x^me^{-\frac{x^2}{2\sigma _{t}^{2}}}dx}}
		\\
		&=\sum_{m=0}^{+\infty}{\left( m+1 \right) c_{m+1}\left[ x_{LO} \right]v_{m}}
	\end{aligned}
\end{equation}
where $v_{m}=v_{m}(x_1-x_{LO},x_2-x_{LO})$ is given by
\begin{equation}
	\begin{aligned}
v_m&=\frac{1}{\sqrt{2\pi}\sigma _t}\int_{x_1-x_{LO}}^{x_2-x_{LO}}{x^me^{-\frac{x^2}{2\sigma _{t}^{2}}}dx}
\\
&=\frac{1}{\sqrt{2\pi}\sigma _t}\left( \int_{x_1-x_{LO}}^{+\infty}{x^me^{-\frac{x^2}{2\sigma _{t}^{2}}}dx}-\int_{x_2-x_{LO}}^{+\infty}{x^me^{-\frac{x^2}{2\sigma _{t}^{2}}}dx} \right) 
\\
&=\frac{2^{\frac{m}{2}-1}\sigma _{t}^{m}}{\sqrt{\pi}}\left( \left( \left( -1 \right) ^m+1 \right) W\left( \frac{m+1}{2} \right) \right. 
\\
&\ \ \ \left. +\left( -1 \right) ^{m+1}W\left( \frac{m+1}{2},\tilde{x}_{1}^{2} \right) -W\left( \frac{m+1}{2},\tilde{x}_{2}^{2} \right) \right) ,
	\end{aligned}
\end{equation}
where $\tilde{x}_{1} = \frac{x_1-x_{LO}}{\sqrt{2}\sigma_{t}}$ and $\tilde{x}_{2} = \frac{x_2-x_{LO}}{\sqrt{2}\sigma_{t}} $. $W(s)$ is gamma function, $W\left( \frac{n+1}{2} \right) =\sqrt{\pi}\frac{\left( n-1 \right) !!}{2^{n/2}}$ and $W\left( s,x \right) =\int_x^{+\infty}{t^{s-1}e^tdt}$ is the incomplete gamma function. 

For the LODC-OFDM system proposed in this work, a symmetrical double clipping operation is implemented at the transmitter, we have $x_{LO}-x_1=x_2-x_{LO}=\Delta x$ and $\tilde{x}_{1}^2=\tilde{x}_{2}^2=\widetilde{\Delta x}^2=\left( \frac{\Delta x}{\sqrt{2}\sigma _t} \right) ^2$. Thus, $v_n$ can be simplified as
\begin{equation}
	\begin{aligned}
v_{2i}=\frac{2^i\sigma _{t}^{2i}}{\sqrt{\pi}}\left( W\left( i+\frac{1}{2} \right) -W\left( i+\frac{1}{2},\widetilde{\Delta x}^2 \right) \right) ,
	\end{aligned}
\end{equation}
where $W \left( \frac{1}{2}+i \right) =\sqrt{\pi}\frac{\left( 2i-1 \right) !!}{2^i}$ and $v_{2i+1}=0$. Finally, by truncating the first $I$ terms, the attenuation factor $\alpha$ can be simplified and the expression shown in Eq. (\ref{alphaeq}).

\section{Nonlinear and Clipping Noise $\sigma_{d}^2$}\label{A3}

Consider $\mathbb{E}\left\{ n_d \right\}=\mathbb{E}\left\{ s_d-\alpha s_t \right\}=\mathbb{E}\left\{ s_d \right\}$ due to $s_t$ follows a zero mean Gaussian distribution, i.e., $\mathbb{E}\left\{ s_t \right\}=0$. We have

\begin{equation}
	\begin{aligned}
\mathbb{E}\left\{ n_d \right\} &=\mathbb{E}\left\{ s_d \right\} 
\\
&=\mathbb{E}\left\{ G\left[ s_t \right] \right\} 
\\
&=\int_{-\infty}^{+\infty}{G\left[ x \right] p\left( x \right) dx}
\\
&=\int_{-\infty}^{+\infty}{F\left[ x+x_{LO} \right]  u\left[ x-\left( x_1-x_{LO} \right) \right]   p\left( x \right) dx}
\\
&\ \  -\int_{-\infty}^{+\infty}{F\left[ x+x_{LO} \right]  u\left[ x-\left( x_2-x_{LO} \right) \right]  p\left( x \right) dx}
\\
&\ \  +F\left[ x_2 \right] \int_{x_2-x_{LO}}^{+\infty}{\frac{1}{\sqrt{2\pi \sigma _{t}^{2}}}e^{-\frac{x^2}{2\sigma _{t}^{2}}}dx}
\\
&\ \  +F\left[ x_1 \right] \int_{-\infty}^{x_1-x_{LO}}{\frac{1}{\sqrt{2\pi \sigma _{t}^{2}}}e^{-\frac{x^2}{2\sigma _{t}^{2}}}dx}
\\
&=\frac{1}{\sqrt{2\pi}\sigma_{t}}\sum_{m=0}^{+\infty}{c_m\left[ x_{LO} \right] \int_{x_1-x_{LO}}^{x_2-x_{LO}}{x^me^{-\frac{x^2}{2\sigma _{t}^{2}}}dx}}
\\
&\ \  +F\left[ x_2 \right] Q\left( \frac{x_2-x_{LO}}{\sigma _{t}} \right) +F\left[ x_1 \right] Q\left( \frac{x_{LO}-x_1}{\sigma _{t}} \right) 
\\
&=\sum_{m=0}^{+\infty}{c_m\left[ x_{LO} \right] v_n}
\\
&\ \  +F\left[ x_2 \right] Q\left( \sqrt{2}\tilde{x}_2  \right) +F\left[ x_1 \right] Q\left( -\sqrt{2}\tilde{x}_1  \right),
	\end{aligned}
\end{equation}
where $Q\left( x \right) =\int_x^{+\infty}{\frac{1}{\sqrt{2\pi}}e^{-t^2/2}dt}$.

For $\mathbb{E}\left\{ s_{d}^{2} \right\} $, the Taylor expansion of $F^2\left[ x+x_{LO} \right]$ is given by
\begin{equation}
	\begin{aligned} 
F^2\left[ x+x_{LO} \right] &=F\left[ x+x_{LO} \right] F\left[ x+x_{LO} \right] 
\\
&=\sum_{m=0}^{+\infty}{\sum_{k=0}^m{c_k\left[ x_{LO} \right] c_{m-k}\left[ x_{LO} \right]}x^m}
\\
&=\sum_{m=0}^{+\infty}{\tilde{c}_m\left[ x_{LO} \right] x^m},
	\end{aligned}
\end{equation}
where $\tilde{c}_m\left[ x_{LO} \right] =\sum_{k=0}^m{c_k\left[ x_{LO} \right] c_{m-k}\left[ x_{LO} \right]}$.

Thus, we have
\begin{equation}
	\begin{aligned} 
\mathbb{E}\left\{ s_{d}^2\right\} &=\mathbb{E}\left\{ G^2\left[ x \right] \right\} 
\\
&=\int_{-\infty}^{+\infty}{G^2\left[ x \right] p\left( x \right) dx}
\\
&=\int_{-\infty}^{+\infty}{F^2\left[ x+x_{LO} \right] u\left[ x-\left( x_1-x_{LO} \right) \right] p\left( x \right) dx}
\\
&\ \  -\int_{-\infty}^{+\infty}{F^2\left[ x+x_{LO} \right] u\left[ x-\left( x_2-x_{LO} \right) \right] p\left( x \right) dx}
\\
&\ \ +F^2\left[ x_2 \right] \int_{x_2-x_{LO}}^{+\infty}{\frac{1}{\sqrt{2\pi \sigma _{t}^{2}}}e^{-\frac{x^2}{2\sigma _{t}^{2}}}dx}
\\
&\ \ +F^2\left[ x_1 \right] \int_{-\infty}^{x_1-x_{LO}}{\frac{1}{\sqrt{2\pi \sigma _{t}^{2}}}e^{-\frac{x^2}{2\sigma _{t}^{2}}}dx}
\\
&=\frac{1}{\sqrt{2\pi}\sigma _t}\sum_{m=0}^{+\infty}{\tilde{c}_m\left[ x_{LO} \right]\int_{x_1-x_{LO}}^{x_2-x_{LO}}{x^me^{-\frac{x^2}{2\sigma _{t}^{2}}}dx}}
\\
& \ \  +F^2\left[ x_2 \right] Q\left( \frac{x_2-x_{LO}}{\sigma _t} \right) 
\\
& \ \ +F^2\left[ x_1 \right] Q\left( -\frac{x_1-x_{LO}}{\sigma _t} \right)
\\
&=\sum_{m=0}^{+\infty}{\tilde{c}_m\left[ x_{LO} \right]v_m}
\\
& \ \  +F^2\left[ x_2 \right] Q\left( \sqrt{2}\tilde{x}_2 \right) +F^2\left[ x_1 \right] Q\left( -\sqrt{2}\tilde{x}_1  \right)  
	\end{aligned}
\end{equation}

For the LODC-OFDM system proposed in this work, a symmetrical double clipping operation is implemented at the transmitter, we have $\tilde{x}_{1}^2=\tilde{x}_{2}^2=\widetilde{\Delta x}^2=\left( \frac{\Delta x}{\sqrt{2}\sigma _t} \right) ^2$ and $v_{2i+1}=0$. By truncating the first $I$ terms, the attenuation factor $\alpha$ can be simplified and the expression shown in Eq. (\ref{noieseq}).

\ifCLASSOPTIONcaptionsoff
  \newpage
\fi



%

\normalem
\bibliographystyle{IEEEtran}
\bibliography{myref}





%

%
%
%




\end{document}